\documentstyle[aps,prd,preprint,epsf,psfig]{revtex}
\tighten
\def\etal{{\it et al.}}
\begin{document}
\draft

\title{Dynamics of Scalar Fields in the Background of\\
       Rotating Black Holes}

\author{William Krivan,${}^{(1,2)}$ Pablo Laguna${}^{(2)}$ 
and Philippos Papadopoulos${}^{(2)}$}

\medskip

\address{
${}^{(1)}$ Institut f\"ur Astronomie und Astrophysik\\ 
            Universit\"at
T\"ubingen, D-72076 T\"ubingen, Germany}

\medskip

\address{
${}^{(2)}$ Department of Astronomy \& Astrophysics and \\
Center for Gravitational Physics \& Geometry\\ 
Penn State University, University Park, PA 16802}

\date{\today}

\maketitle

\begin{abstract}
A numerical study of the 
evolution of a massless scalar field in the background of 
rotating black holes is presented.
First, solutions to the wave equation are obtained for
slowly rotating black holes. In this approximation,
the background geometry is treated
as a perturbed Schwarzschild spacetime with the angular momentum
per unit mass 
playing the role of a perturbative parameter.
To first order in the angular momentum of the black hole, the 
scalar wave equation
yields two coupled one-dimensional evolution equations for a 
function
representing the scalar field in the Schwarzschild background 
and a 
second field that accounts for the rotation. 
Solutions to the wave equation are also obtained for rapidly 
rotating
black holes. In this case, the wave equation does not admit
complete separation of variables and yields a two-dimensional 
evolution equation.
The study shows that, for rotating black holes, the
late time dynamics of a massless scalar field exhibit the same 
power-law behavior
as in the case of a Schwarzschild background independently of
the angular momentum of the black hole.
\end{abstract}


\section{Introduction}
\label{intro}

Perhaps the simplest wave phenomenon in relativity is
the propagation of linearized waves on a fixed, curved 
background.
When observed at a fixed spatial point,
the dynamics of a wave propagating on spherically symmetric, 
time-independent, asymptotically flat, background geometries
consists of three stages:
During the first stage, the observed wave depends on the
structure of the initial pulse and its reflection from the 
origin (Burst phase). This phase is followed by 
an exponentially decaying quasi-normal ringing of the
black hole (Quasi-normal phase). In the last stage,
the wave slowly dies off as a power-law tail (Tail phase).
The last two phases are dictated by the interference of the part
of 
the wave backscattered at the tail of
the potential and the part reflected by the potential barrier.

The tail phenomenon can be understood as due to the scattering
of the wave off the effective curvature potential of the
background geometry \cite{price72,cpm}. Tails have been mostly 
investigated  
on Schwarzschild backgrounds \cite{waimo95a,waimo95b,gpp94a},
Reissner-Nordstrom black-holes \cite{gpp94a},
and for collapsing
scalar fields\cite{gomez92,gpp94b}.
The study of tails has also implications in connection with the
gravitational radiation emerging from inspiraling 
binary systems\cite{apostolatos93}.

For spherically symmetric systems, after separation of variables,
the three-dimensional wave equation $\nabla_\mu\nabla^\mu\Phi = 
0$
reduces, in a suitable radial coordinate $x$, to the 
one-dimensional wave equation:
\begin{equation}
    [-\partial^2_t + \partial^2_x - V(x)]\Phi(x,t) = 0 \, ,
\label{eq:kg}
\end{equation}
where $V(x)$ denotes the effective curvature potential.
Recently, the late-time behavior of equation (\ref{eq:kg})
has been the subject of detailed numerical and analytic
investigations by Gundlach \etal
\cite{gpp94a} and Ching \etal \cite{waimo95a,waimo95b}.

Gundlach \etal \cite{gpp94a,gpp94b} studied both, the late-time
behavior of massless fields around a fixed Schwarzschild 
geometry and
the full nonlinear evolution of a minimally coupled scalar
field. Their study shows a remarkable agreement between the
numerically computed quasi-normal frequencies and power-law 
tails and
their corresponding theoretical predictions.  Using a null
characteristic approach, they found tails not only at timelike
infinity, but also at future null infinity and at the future 
horizon
of the black hole. The time behavior of the tails is shown to be

\begin{equation}
\Phi \sim \left\{  
\begin{array}{ll}
t^{-(2l+P+1)}  &\mbox{at timelike infinity,} \\
(t-x)^{-(l+P)}   &\mbox{at future null infinity,} \\
(t+x)^{-(2l+P+1)} &\mbox{at future horizon.} 
\end{array}\right. 
\end{equation}
with $P=1$ if there is an initial static field and
$P=2$ otherwise.

Pursuing a different approach, Ching \etal 
\cite{waimo95a,waimo95b}
showed that the late time tail phenomena are governed by the
asymptotic structure of the background spacetime.  Furthermore, 
they
found that, for non-static initial conditions, the simple 
power-law
$\Phi \sim t^{-2l+\alpha}$ in Schwarzschild geometries, is not 
present
for general potentials of the form $V(x) \sim l(l+ \nolinebreak 
1)/x^2
+ x^{-\alpha}\ln{x}$.  Instead, the generic late-time behavior is
given by $ \Phi \sim t^{-2l+\alpha}\, \ln t$.

In spite of previous work, a complete understanding of the tail
phenomenon for general systems is still missing.  
Questions such as
which factors determine the magnitude and power-law time 
dependence
have not been addressed for a general system.  It is of 
particular
interest, for instance, to show which role the dimensionality 
of the
system plays on the existence of power-law tails\cite{price}.  
When
the point of view that tails arise from backscattering off an
effective potential is adopted, it is not clear whether the tail
behavior would remain unchanged, or be present at all, for
intrinsically two or three dimensional systems \cite{krivan96}.

A natural multi-dimensional generalization of the dynamics of 
scalar
fields is obtained by considering the family of Kerr spacetimes.
In
contrast to the Schwarzschild background geometry, the wave 
equation
$\nabla_\mu\nabla^\mu\Phi = 0$ in the background of a rotating 
black
hole only admits separation of the azimuthal coordinate, so in
principle a two-dimensional evolution problem has to be solved.
It has been suggested\cite{waimo95a}, that tails similar to 
those present
in Schwarzschild spacetimes should exist on Kerr backgrounds, a
detailed analysis, however, has not been undertaken.

This study considers the dynamics of scalar fields on the 
background
of rotating black holes in two regimes: slowly and rapidly 
rotating
holes.  A considerable mathematical simplification of the 
problem is
achieved in the case of slowly rotating black holes since the 
angular
momentum can be treated as a perturbative parameter.  Thus, to 
first
order in the angular momentum, the scalar wave equation yields 
two
one-dimensional evolution equations.  This formalism is 
described in
Section \ref{formal}.  Results of the numerical calculations are
presented in Section \ref{results}, where power-law tails are
discussed.  The case of rapidly rotating black holes is 
discussed in
Section \ref{rapid}.

\section{Slowly Rotating Black Holes}
\label{formal}

Using Boyer-Lindquist coordinates ($t,r,\theta,\phi$) the wave 
equation for a massless scalar field $\Phi$ reads \cite{brill72}
\begin{eqnarray}
\label{BL}
&&-\left[ \frac{(r^2+a^2)^2}{\Delta} - a^2 \sin^2\theta \right]
\,
\partial_{tt} \Phi - \frac{4Mar}{\Delta}\, \partial_{t\phi} \Phi
+\left[\frac{1}{\sin^2\theta}-\frac{a^2}{\Delta}\right]
\partial_{\phi\phi}  \Phi \nonumber \\
&&+\partial_r\left(\Delta \partial_r\Phi\right) 
+\frac{1}{\sin\theta}\partial_\theta\left(
\sin\theta\partial_\theta\Phi\right)
=0 \, ,
\end{eqnarray}
where $M$ is the mass of the black hole,
$a$ is its angular momentum per unit mass, and $\Delta\equiv
r^2-2Mr+a^2$. Equation (\ref{BL}) is equivalent to the special
case of the Teukolsky equation \cite{teuk73} for fields with 
vanishing spin weight.

The Schwarzschild case, $a=0$, allows separation of
variables in terms of the scalar spherical harmonics 
$Y^m_l(\theta,\phi)$
without any requirements on a particular time behavior.
The solution of (\ref{BL}) can be written in the form
\begin{equation}
\label{spheresol}
\Phi =\frac{1}{r}\,Y^m_l(\theta,\phi)\,\Psi(t,r^*)\, ,
\end{equation}
where $r^*$ denotes the tortoise coordinate 
\begin{equation}
\label{tort}
r^* \equiv r+2M\ln(r-2M) \; .
\end{equation}
Substitution of (\ref{spheresol}) into (\ref{BL}), with $a=0$,
yields the one-dimensional wave equation
\begin{equation}
\label{sphereeq}
[-\partial_{tt} +\partial_{r^*r^*} - V(r) ] \,\Psi(t,r^*) = 0\;,
\end{equation}
with the potential $V(r)$ defined by
\begin{equation}
\label{sspot}
V(r)= \left(1-\frac{2M}{r}\right)\left[\frac{l(l+1)}{r^2} +
\frac{2M}{r^3}\right] \; .
\end{equation}
It has been shown, in both numerical and analytic studies
\cite{price72,waimo95a,gpp94a}, that at a fixed radius
the solution of (\ref{sphereeq}) will fall off as $t^{-(2l+3)}$ 
for large $t$ and non-static initial data.

For $a \ne 0$, it is no longer possible to separate variables and
arrive at an equation similar to (\ref{sphereeq}).  While the
azimuthal dependence of $\Phi$ can still be described by $e^{i
m\phi}$, a full separation of variables exists only if the time
dependence is given by $e^{i\omega t}$, as first demonstrated by
Brill \etal \cite{brill72}.  In this case, both, the Teukolsky 
equation and
equation (\ref{BL}), admit separable solutions of the form
\begin{equation}
\Phi = e^{-i\omega
t}e^{-im\phi}\,_sS^m_l(\theta;\omega)\,R_{lm}(r;w)\, ,
\end{equation}
where $\,_s S^m_l$ are the spin-weight-$s$ spheroidal harmonics.
However, since the objective of the present study is the 
late-time
dynamics, a decomposition based on $\omega$ modes is not 
suitable, and
one is forced, in principle, to solve a two-dimensional evolution
equation for $\Phi$.  Nonetheless, the case of slowly rotating 
black
holes circumvents the problem of solving a two-dimensional wave
equation.

The use of an azimuthal coordinate defined by asymptotic 
observers, as
is the case with the Boyer-Lindquist $\phi$ coordinate, 
introduces
unphysical pathologies near the horizon even in the case $a<<M$.
A discussion of those pathologies and their precise 
manifestation in
the slow rotation limit is given in the Appendix.
Those coordinate induced problems are readily dealt with by
adopting the Kerr azimuthal coordinate $\tilde \phi$ given by
\begin{equation}
\label{kerrphi}
	d\tilde\phi=d\phi+\frac{a}{\Delta}\, dr \, .
\end{equation}
In the ($t,r,\theta,\tilde\phi$) coordinates, equation~(\ref{BL})
is transformed into
\begin{eqnarray}
\label{teuks0}
&&-\left[ \frac{(r^2+a^2)^2}{\Delta} - a^2 \sin^2\theta \right]
 \,
\partial_{tt} \Phi - \frac{4Mar}{\Delta}\, 
\partial_{t\tilde\phi} \Phi
+ 2a  \partial_{r\tilde\phi} \Phi
+\partial_r\left(\Delta \partial_r\Phi\right)  \nonumber \\
&&+\frac{1}{\sin\theta}\partial_\theta\left(
\sin\theta\partial_\theta\Phi
\right)
+\frac{1}{\sin^2\theta} \partial_{\tilde\phi\tilde\phi} \Phi
=0 \, ,
\end{eqnarray}

Under the assumption $a<<M$, one can seek a solution of 
(\ref{teuks0})
that possesses an angular dependence given by
$Y_l^m(\theta,\tilde\phi)$ following a procedure similar to that
used
in deriving (\ref{spheresol}).  That is, one views the function
$\Psi(t,r^*)$ containing a piece $\Psi_{0}(t,r^*)$ representing 
the
scalar field in the Schwarzschild background and a second field
$\Psi_{1}(t,r^*)$ that takes into account the correction due to 
the
rotation of the black hole to first order in $a$.  Thus, the 
ansatz
for the solution of (\ref{teuks0}) has the form
\begin{equation}
\label{ansatz}
\Phi
=\frac{1}{r}\,Y^m_l(\theta,\tilde\phi)\,[\Psi_{0}(t,r^*)+a\,
\Psi_{1}(t,r^*)]
\; .
\end{equation}

Before deriving the equations satisfied by $\Psi_{0}$ and 
$\Psi_{1}$,
a further transformation of (\ref{teuks0}) to the coordinate 
system
$(t,r^*,\mu,\tilde\phi)$ is performed, where $r^*$ denotes the
Schwarzschild tortoise coordinate (\ref{tort}) and $\mu \equiv
\cos\theta.$ In these coordinates, substitution of (\ref{ansatz})
into the wave equation $\nabla_\mu\nabla^\mu\Phi = 0$ and 
collecting
powers of $a$ yields after separation of variables the equation
\begin{equation}
\label{taylor1}
  \Box_{0} \Psi_0 + a \left( \Box_{0} \Psi_1 - \rho_0 
\right) +
{\cal{O}}(a^2) = 0 \, ,
\end{equation}
where
\begin{mathletters}
\begin{eqnarray}
\label{box0}
\Box_{0} &\equiv& -\partial_{tt} +\partial_{r^*r^*} - V(r) \quad 
\mbox{and} \\ 
\label{source}
\rho_0 &\equiv&  - 2\,i\,m\,\frac{-2Mr\partial_t\Psi_{0}
	   +r^2\partial_{r^*}\Psi_{0}
	   -(r-2M)\Psi_{0}}{r^4} \, .
\end{eqnarray}
\end{mathletters}

The system of equations for $\Psi_{0}$ and $\Psi_{1}$ is 
obtained by
requiring that the zero and first order terms of this expansion 
vanish
independently, namely
\begin{mathletters}
\label{numsys}
\begin{eqnarray}
\left[ -\partial_{tt} + \partial_{r^*r^*} - V(r) \right] \, 
\Psi_{0} &=& 0
\; , \label{numsys0} \\
\left[ - \partial_{tt} + \partial_{r^*r^*} + V(r) \right] \, 
\Psi_{1}
&=& 2\,i\,m\,\frac{-2Mr\partial_t\Psi_{0}
	   +r^2\partial_{r^*}\Psi_{0}
	   -(r-2M)\Psi_{0}}{r^4}
\label{numsys1}
\end{eqnarray}
\end{mathletters}
with $V(r)$ given by (\ref{sspot}).
As expected, the equation for $\Psi_{0}$ does not contain terms
depending on $\Psi_{1}$, and the source term $\rho_{0}$ 
in the
equation for $\Psi_{1}$ only depends on the solution $\Psi_0$ 
of the
zero order equation.  The problem has then been reduced to 
solving the
homogeneous equation (\ref{numsys0}) for $\Psi_{0}$ and using 
this
solution as a source in the inhomogeneous equation 
(\ref{numsys1}).

Owing to the one-way membrane character of the horizon, the use 
of
ingoing boundary conditions at the horizon arises most 
naturally for
the solution of (\ref{numsys0}).  The asymptotic form of 
$\Psi_{0}$
for $r^* \rightarrow - \infty$ is then a wave with constant 
amplitude
propagating to the horizon.  

As $r\rightarrow 2M$ ( $r^* \rightarrow - \infty$ ), the source 
term
behaves as $ \rho_0 \rightarrow
-\partial_t\Psi_{0}+\partial_{r^*}\Psi_{0}.$ Thus, $ \rho_0 
\rightarrow
0$ if $\Psi_{0} \rightarrow \Psi_{0}(t+r^*)$.  On the other 
hand, at
$r^* \rightarrow \infty$, $ \rho_0 \rightarrow 0$ due to 
the
assumption that $\Psi_{0}$ has compact initial data.

\section{Power-Law Tails}
\label{results}
The numerical results presented here were computed with a second 
order staggered in time evolution scheme.
For consistency, a null characteristic numerical integration 
was also used, and the results were in complete agreement.

In general, both $\Psi_{0}$ and $\Psi_{1}$ are complex
quantities. Owing to the purely imaginary coefficients 
of $\partial_t\Psi_{0}$, $\partial_{r^*}\Psi_{0}$, and 
$\Psi_{0}$ in the source term in (\ref{numsys1}), 
$\mbox{Im}\Psi_{0}$ is coupled to $\mbox{Re}\Psi_{1}$, and 
$\mbox{Re}\Psi_{0}$ to $\mbox{Im}\Psi_{1}$.  
Hence, without loss of generality, one may assume 
$\mbox{Re}\Psi_{0}\equiv
\mbox{Im}\Psi_{1} \equiv 0$.
Since the separation variable $m$ plays the
role of a scaling factor for the source term in (\ref{numsys1}), 
in the numerical calculations $m=M=1$ is used. 

The initial data for $\Psi_{0}$ consist of
a bell-shaped pulse propagating outwards given by 
\begin{equation}
\label{inidata}
 \Psi_{0}^{\mbox{\scriptsize ini}} = c_1 
\left[ ( c_2(r^*-t) - r^*_{\mbox{\scriptsize in}})( c_2(r^*-t) -
r^*_{\mbox{\scriptsize out}}) \right]^8
\end{equation}
for $r_{\mbox{\scriptsize in}} \leq r^* \leq r_{\mbox{\scriptsize
out}}$, and vanishing otherwise.  The parameters have been 
chosen to
yield a pulse centered at $r^*=100$; that is, $c_1 = 100, 
c_2 = 0.02$
and $r^*_{\mbox{\scriptsize in,out}} = 1,3$.  On the other hand,
the
initial conditions for $\Psi_{1}$ are $\Psi_{1}= \partial_t
\Psi_{1}=0$.
Numerical experiments have shown, that the tail behavior is not
affected by the choice of initial conditions for $\Psi_{1}$.

As previously discussed, the behavior of both the potential and 
the
source term in (\ref{numsys1}) makes it possible, in principle, 
to
impose ingoing boundary conditions at the horizon and outgoing
conditions at infinity, i.e.\ $\lim_{r^* \to
\mp\infty}\Psi_{0,1}=\Psi_{0,1}(t\pm r^*)$.  However, in a Cauchy
evolution, such as the one here under consideration, boundary
conditions are imposed at a finite distance \cite{PL}.  
Typically, the
computational domains used covered $-50 \leq r^* \leq 2000$ with
$\Delta r^* = 0.1$.  Since $r^* = -50$ yields $r - 2M 
\sim 10^{-12}$,
imposing the left (ingoing into the black hole) boundary 
condition at
a finite radius $r^*$ turns out to be a suitable approximation; 
both,
the potential $V(r)$ and the source $\rho_{0}$ are 
negligible at $r^*
= -50$. A different situation is encountered at the right 
boundary.
Even though the potential $V(r)$ and the source $\rho_{0}$
vanish as
$r^* \rightarrow \infty$, at the right boundary ($r^* = 2000$) 
of the
computational domain, and especially for late times, the outgoing
boundary condition is not a good approximation.  However, this
boundary is at a sufficiently large radial distance, so it 
allows enough
dynamical evolution range to obtain the tail behavior before 
numerical
boundary effects contaminate the solution.
 
\begin{figure}[h]
\leavevmode
\\
\epsfxsize=0.5\textwidth
\epsfbox{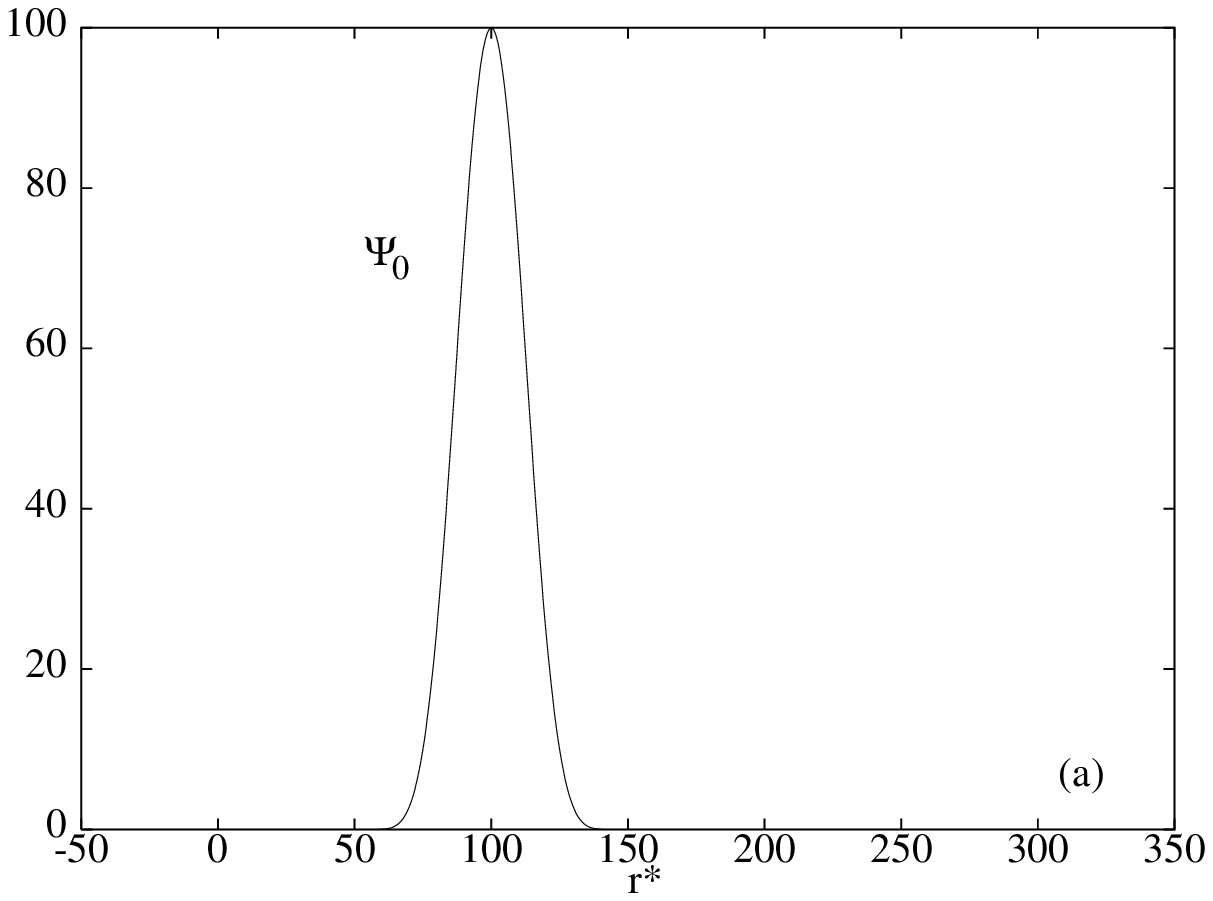} 
\epsfxsize=0.5\textwidth
\epsfbox{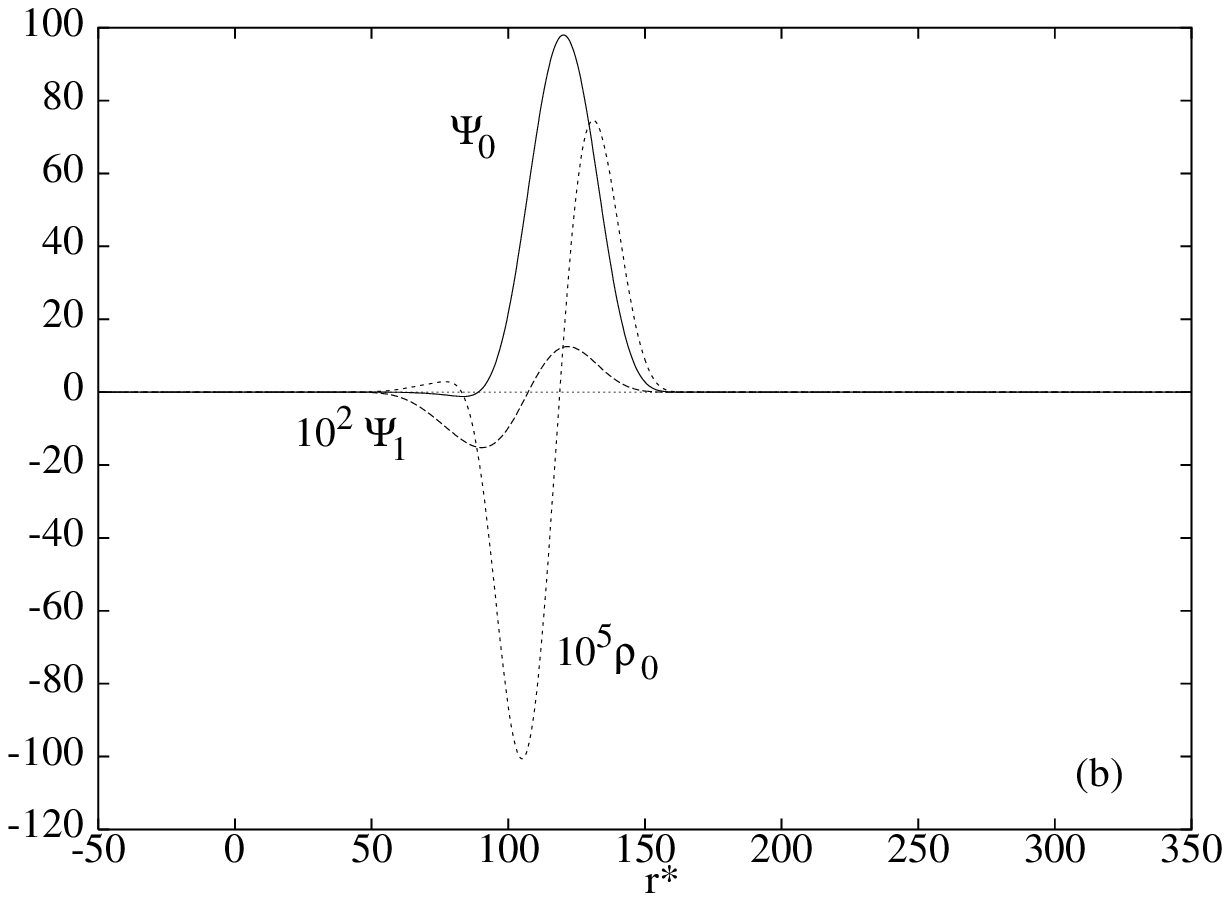}\\
\epsfxsize=0.5\textwidth
\epsfbox{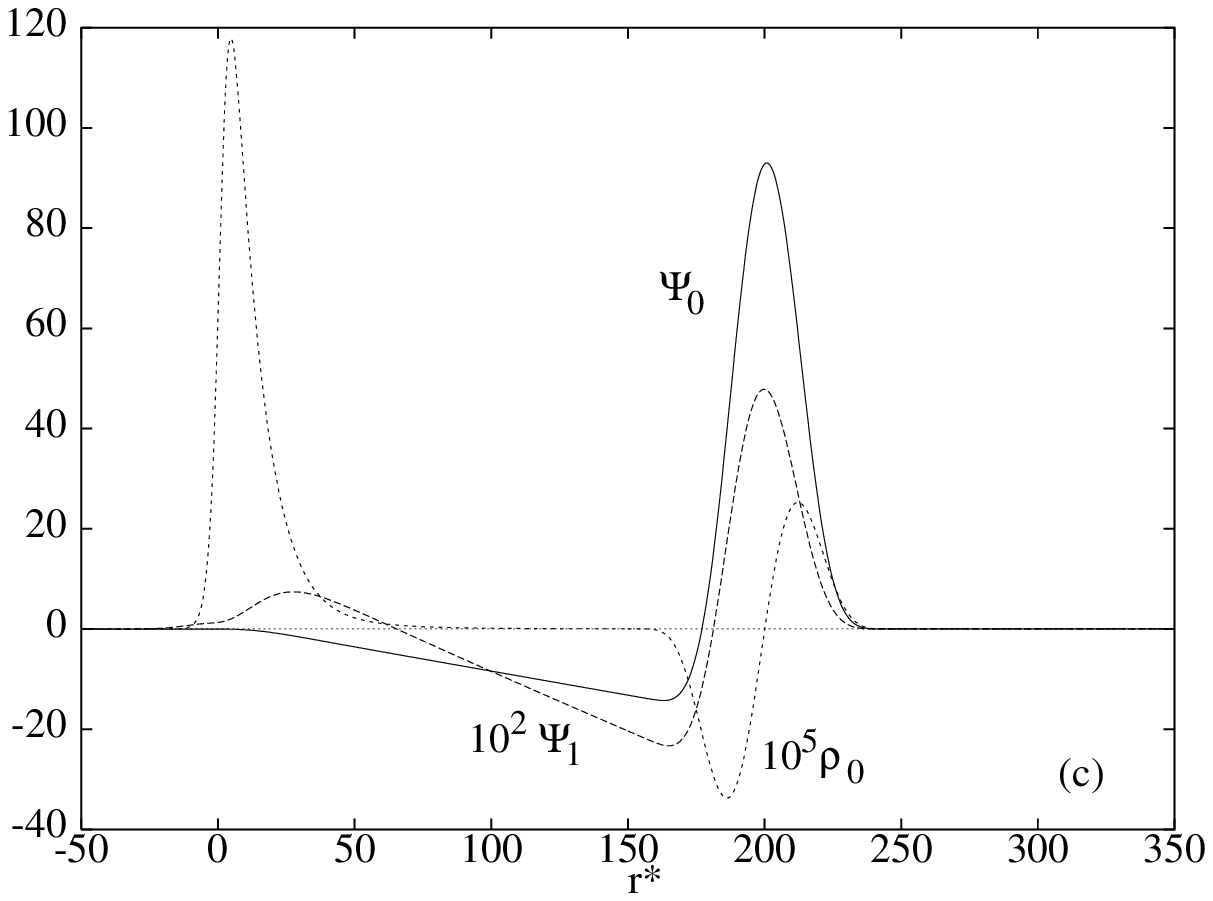}
\epsfxsize=0.5\textwidth
\epsfbox{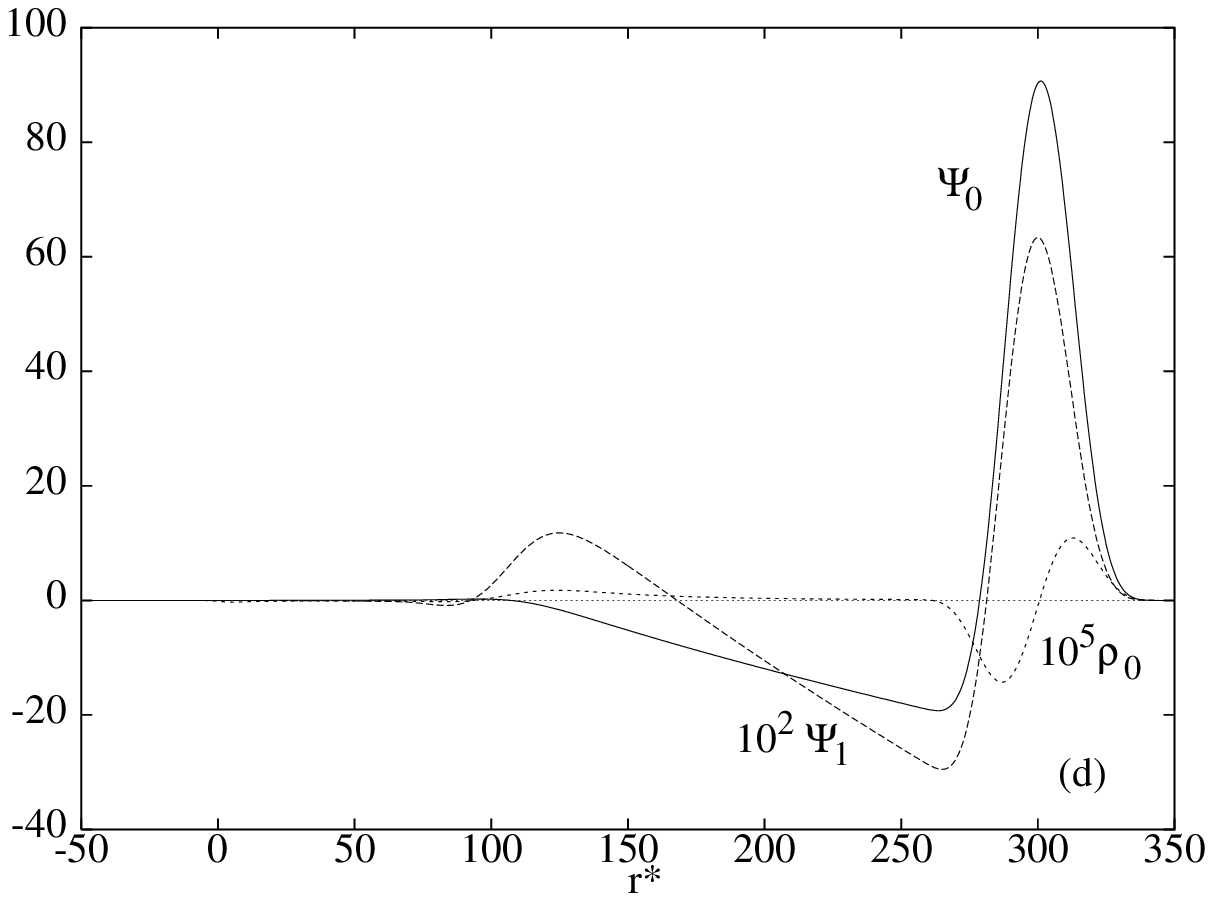}
\caption[figure1]{\label{figure1}
Evolution of $\Psi_{0}$, $ \rho_0$, and $\Psi_{1}$ for
for $l=1$, $0\leq t \leq 200$. (a) At $t=0$, $\Psi_{1}=
\rho_0=0$. (b) At $t=20$, the
source $ \rho_0$ is dominated by the initial shape of
$\Psi_{0}$. (c) At $t=100$,
a peak in $ \rho_0$ has developed due to the part of
the initial pulse that was backscattered off the potential and is
propagating inwards. (d) At $t=200$, 
as a consequence of the peak of $ \rho_0$ in (c), a hump 
appears in the
trailing part of $\Psi_{1}$. }
\end{figure}

Figure\ \ref{figure1} shows several snapshots of the evolution 
for
$l=1$, $0\leq t \leq 200$.
During the evolution, the initially outgoing
zero order function $\Psi_{0}$ is followed by a pulse in
$\Psi_{1}$ driven by the source $ \rho_0$.  
In Fig.\ 1(a), the initial bell-shaped pulse for $\Psi_0$ 
centered at
$r^*=100$ is shown. $\Psi_1=\rho_0=0$. 
At $t=20$, Fig.\ 1(b) depicts the radial dependence of the source
$ \rho_0$ dominated by the initial shape of $\Psi_{0}$. As
displayed in Fig.\ 1(c),
after an evolution time of $t=100$,
a pronounced peak of $ \rho_0$ has developed due to the 
part of
the initial pulse that was backscattered off the potential and is
propagating inwards. In Fig.\ 1(d), at $t=200$, 
as a consequence of the peak of $ \rho_0$, a hump appears 
in the
trailing part of $\Psi_{1}$.

Before presenting the numerical results, a heuristic analytical
argument is given estimating the expected late time behavior.  
The
argument is based on the result that the late time decay of waves
propagating on curved spacetimes is dictated by the spatial
asymptotics of the potential \cite{waimo95b}.  The argument does
 not
include the contribution from the centrifugal barrier ($l=0$).
Starting with the original wave equation (\ref{eq:kg})
one obtains to first order in $a$ 
\begin{equation}
\label{1stordteuks0}
\left[ -\partial_{tt} + \partial_{r^*r^*} - V(r) \right]\,\Psi 
  + \frac{2\,i\,m\,a\,M}{r^3} 
   \left[  - 2\,\partial_t 
  + r\, \partial_{r^*}
  + \left( 1 - \frac{2M}{r} \right)  \right] \Psi
  =0 \, ,
\end{equation}
where
\begin{displaymath}
\Phi = \Psi(t,r^*) Y^m_l(\theta,\tilde\phi).
\end{displaymath}
Equation (\ref{1stordteuks0}) consists of two operators: the 
operator
for scalar fields on Schwarzschild background
(first term in squared brackets) and a second operator
(first order in $a$)
containing first order spatial and temporal derivatives. 

If one views the late-time tail behavior at a radial position 
$r$,
as the result of the scattering by a potential at $\hat r >> r$ 
of
a wave originated at $r_o \sim r$, the arrival time of the
scattered wave is approximately given by
$t \approx (\hat r-r_o)+(\hat r-r) \approx 2\hat r$
\cite{waimo95a,waimo95b}.
At late times, the scalar field $\Psi \propto V_T(\hat r) 
\approx V_T(t)$,
where $V_T(r,t) = V(r) + V_a(r,t)$ represents the ``total" 
potential; that is,
the Schwarzschild potential (\ref{sspot}) with $l=0$ plus a 
correction
$V_a$ due to the rotation of the black hole.
This correction arises from the second term in squared brackets 
in 
(\ref{1stordteuks0}).
Under the late-time and large scattering radius approximation,
$\partial_t\Psi \sim \Psi/t$ and $\partial_{r^*}\Psi \sim 
\Psi/r$.
Thus, $V_a(\hat r,t) \sim \hat r^{-3} + \hat r^{-3}\,t^{-1} + 
\hat r^{-4} \sim t^{-3}$.
Therefore, $\Psi \propto V + V_a \sim t^{-3}$ since $V \sim 
t^{-3}$.
This shows that both, the Schwarzschild potential
and the perturbative potential, contribute with the same power 
law behavior.
The centrifugal barrier adds a $t^{-2\,l}$ factor to the tails 
\cite{waimo95b}.

\begin{figure}[h]
\leavevmode
\epsfxsize=0.85\textwidth
\epsfbox{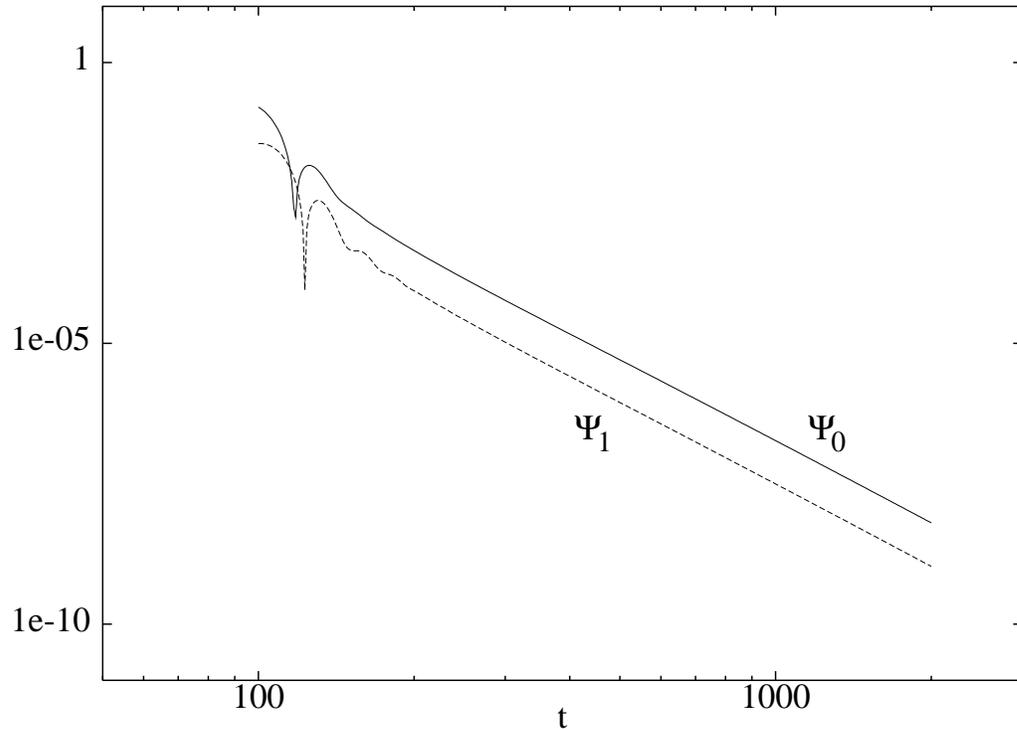}
\caption[figure2]{\label{figure2}
Log-log plots of $\Psi_{0}$ and $\Psi_{1}$ for $l=1$ at $r^* =
10$. The power-law exponents are $-4.93$ for $\Psi_{0}$ and 
$-4.92$
for $\Psi_{1}$. The wiggles that can be seen for $t<200$ are
remainders of the quasi-normal ringing of the black hole.
}
\end{figure}

\begin{figure}[h]
\leavevmode
\epsfxsize=0.85\textwidth
\epsfbox{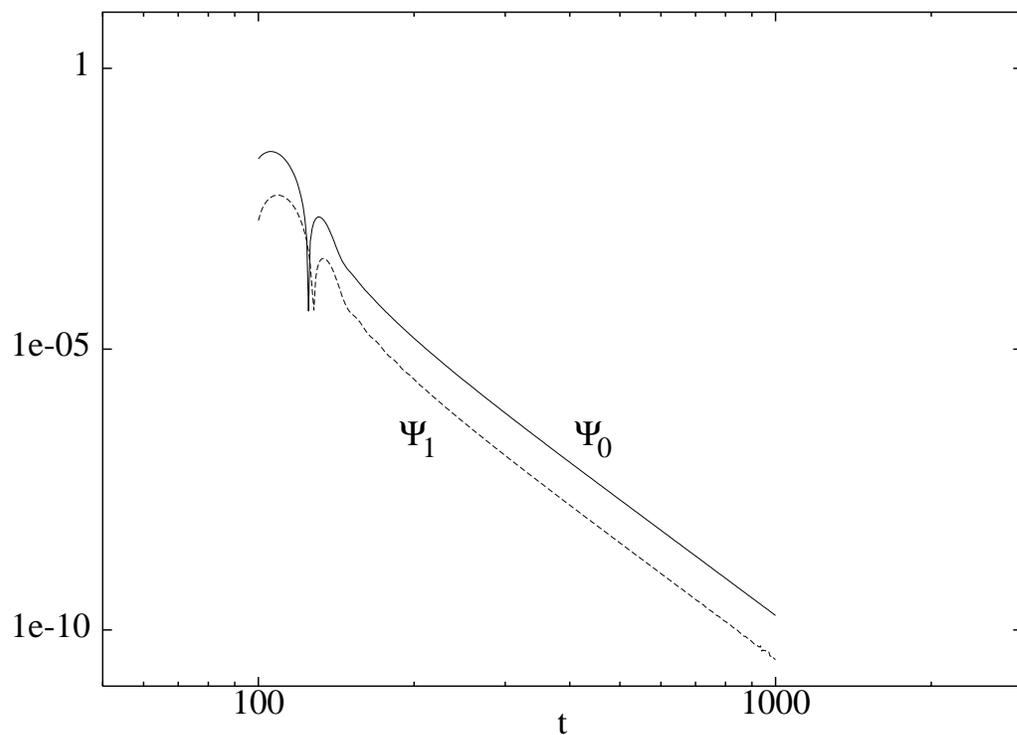}
\caption[figure3]{\label{figure3}
Log-log plots of $\Psi_{0}$ and $\Psi_{1}$ for $l=2$ at $r^* =
10$. The power-law exponents are given by $-7.02$ and $-7.07$, 
respectively.
}
\end{figure}

The numerical investigation of the late time behavior was 
performed 
at different distances from the black hole. Power-law behavior, 
with
an exponent independent of the location of observation, was 
detected 
for both $\Psi_{0}$ and $\Psi_{1}$ for different multipole 
indices $l$. 
As shown in Fig.\ \ref{figure2} and Fig.\ \ref{figure3}, the
numerically determined power-law
exponents for the zero and first order solutions are in good
agreement: For $l=1$, the power-law exponents are $-4.93$ for
$\Psi_{0}$ and $-4.92$ for $\Psi_{1}$; the theoretically
predicted value is 5 for this case. For $l=2$, the exponents are
$-7.02$ and $-7.07$, respectively, and the corresponding 
theoretical
value is 7.
These results lead to a power-law exponent for 
$\Phi=\frac{1}{r}\,Y^m_l \,[\Psi_{0}+a\,\Psi_{1}]$
that is unchanged with respect to the behavior on a 
Schwarzschild background.
\section{Rapidly Rotating Black Holes}
\label{rapid}

The next step is to consider the case of rotating black holes 
with arbitrary angular momentum per unit mass, i.\ e.\ $0\leq a 
\leq M$.
The objective here is to investigate whether the Schwarzschild 
tail 
behavior is modified
for rapidly rotating black holes.
The starting point is equation (\ref{teuks0}) written in 
coordinates
$(t,r^*,\theta,\tilde\phi)$, with the Kerr tortoise coordinate 
$r^*$
defined by
\begin{equation}
\label{kerrtort}
\frac{dr^*}{dr} \equiv \frac{r^2+a^2}{\Delta} \, .
\end{equation}

Using the ansatz
\begin{equation}
\label{kerransatz}
\Phi \equiv \Psi(t,r^*,\theta)\,e^{im\tilde\phi},
\end{equation}
the wave equation (\ref{teuks0}) yields
\begin{eqnarray}
\label{2Dsep}
&&-\partial_{tt} \Psi - \frac{(r^2+a^2)^2}{\sigma}\, 
\partial_{r^*r^*} \Psi
-\frac{\Delta}{\sigma}\, \partial_{\theta\theta} \Psi
+\frac{4\,i\,m\,a\,r\,M}{\sigma}\,\partial_t \Psi
-\frac{2\,\left[r\Delta+i\,a\,m\,(r^2+a^2)\right]}{\sigma}\,
\partial_{r*}
\Psi  \nonumber \\
&&-\frac{\cos\theta\,\Delta}{\sin\theta\,\sigma}\,
\partial_{\theta} \Psi
+\frac{m^2\,\Delta}{\sin^2\theta \, \sigma}\, \Psi \, =0  ,
\end{eqnarray}
with
\begin{equation}
\label{sigmadef}
\sigma \equiv
-r^4-r^2a^2\,(1+\cos^2\theta)-2a^2rM\sin^2\theta-a^4\cos^2\theta
\, .
\end{equation}
The wave equation (\ref{2Dsep}) reduces to the flat wave
equation $[-\partial_{tt}+\partial_{r^*r^*}]\Psi=0$ for $r^* 
\rightarrow
\pm \infty$, thus a simple formulation of
radial outgoing boundary conditions is possible.

Equation (\ref{2Dsep}) was evolved on a rectangular spatial grid
in 
$r_{\mbox{\scriptsize min}} \leq r \leq r_{\mbox{\scriptsize 
max}}$, $0
\leq \theta \leq  \pi/2 $. 
The numerical code was tested for convergence and stability.  
As in
the one-dimensional case, the mass of the black hole was set to 
$M=1$.
The inner radial boundary $r_{\mbox{\scriptsize min}}$ was 
chosen so
that the limit of the flat wave equation was reasonably 
approximated
(cf.\ Section \ref{results}).  The data at $\theta=0,\pi/2$ were
updated according to the behavior of the particular mode under
consideration.  The values of the field at the poles and the
equatorial plane were determined by the value of $m$ and the 
symmetry
of the mode with respect to the equatorial plane: For $m=0$,
$\partial_\theta\Psi(t,r^*,0) =0$ for all $t,r^*$. For
non-axisymmetric waves, given by $m\neq0$, $\Psi(0)=0$ for all 
modes.
At the equator, the condition for the modes symmetric to the 
equatorial
plane is given by $\partial_\theta\Psi(\pi/2)=0$ and the 
antisymmetric
modes are characterized by $\Psi(\pi/2)=0$.

The initial data were given by 
$\mbox{Re}\Psi = \partial_t \mbox{Re}\Psi= 0$ and, using 
(\ref{inidata}),
$\mbox{Im}\Psi = \Psi_{0}^{\mbox{\scriptsize ini}} \, P_l^m / 
\sqrt{r^2+a^2}$, 
where a particular associated Legendre Polynomial $P_l^m$ had 
to be
chosen according to the mode of interest. Because $l$ does not 
appear
in the equation, for fixed $m$, a mixing of modes belonging
to different values of $l$ will occur during the evolution if the
initial data do not correspond to the lowest possible mode.

\begin{figure}[h]
\leavevmode
\epsfxsize=0.85\textwidth
\epsfbox{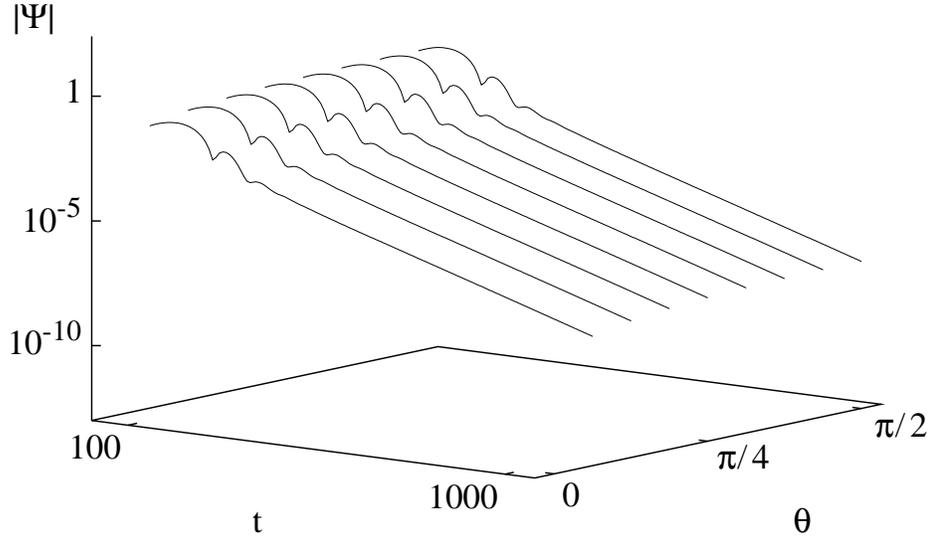}
\caption[figure4]{\label{figure4}
                 Log-log plots of $|\Psi |$ for $m=1$, $a=0.99$
                 at  $r^* = 10$ for different equidistant 
angular directions
                 from $\theta = \pi/16$ to
                 $\theta = \pi/2$.
                 The initial angular dependence was given by
                 $\mbox{Im}\Psi \sim P_1^1\sim \sin\theta$. 
For large times
                 the time dependence is given by
                 $|\Psi | \sim t^{-\mu}$, where $\mu = 4.87\pm0.
005$ for all
                 observed angles. The theoretical power-law tail
exponent in the Schwarzschild case is given by $\mu=5$. 
                       }
\end{figure}

\begin{figure}[h]
\leavevmode
\epsfxsize=0.85\textwidth
\epsfbox{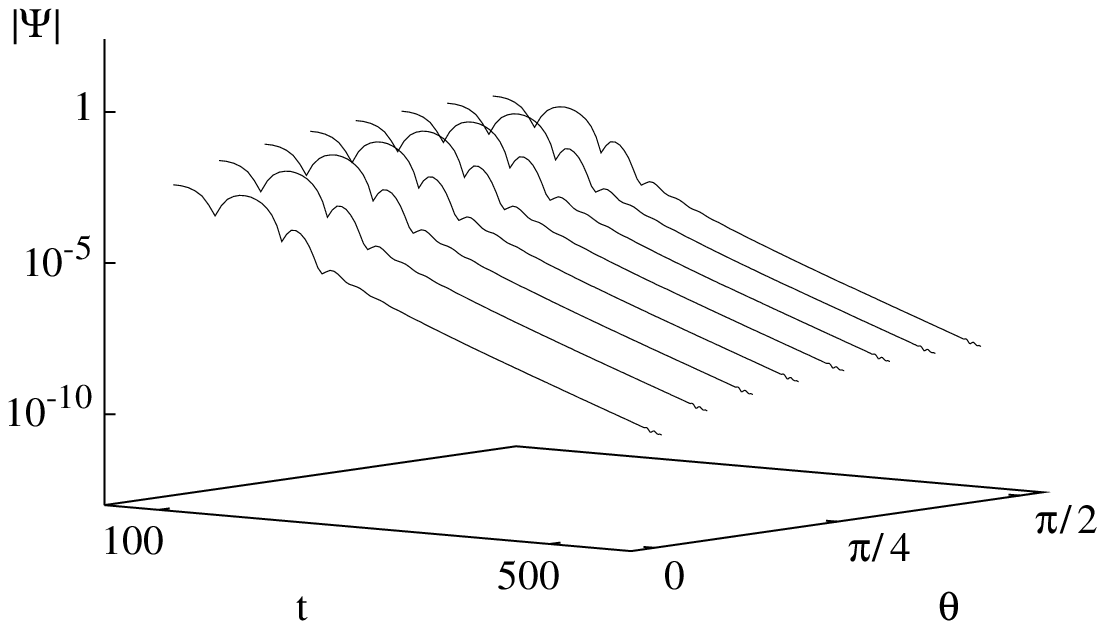}
\caption[figure5]{\label{figure5}
                 Same as in Fig.\ 4, but for $m=2$ and 
                 the initial angular dependence given by
                 $\mbox{Im}\Psi \sim P_2^2 \sim \sin^2\theta$.
                 For large times the power-law exponent
                 is $\mu =6.98\pm 0.004$ for all observed
                 angles. The 
                 theoretical value
                 in the Schwarzschild case is given by $\mu=7$.
                       }
\end{figure}

In the following figures, 
the late-time behavior of $| \Psi |$ at $r^* = 10$ is
displayed for different initial multipoles for an angular 
momentum
parameter of $a=0.99$.
Fig.\ \ref{figure4} shows the late time behavior for $m=1$.
Equidistant angular directions were chosen in the interval 
from $\theta = \pi/16$ to $\theta = \pi/2$. The initial angular
dependence corresponds to $l=m=1$. The late-time behavior can be
described by $| \Psi | \sim t^{-\mu}$, where the 
exponent is given by $\mu = 4.87 $ for all the displayed angles.
In Fig.\ \ref{figure5} the analogous situation is shown for 
$m=2$ and an 
initial pulse corresponding to $l=m=2$. In this case the
power-law exponent for the observed angles is $\mu=6.98$.
That is, the power-law exponents governing the late-time 
behavior for
$a=0.99$ do not exhibit a significant change when compared to the
Schwarzschild case, in which
the theoretical power-law tail exponents are given by $\mu=5$ for
$l=m=1$, and $\mu=7$ for $l=m=2$. 
In both cases the initial pulse used for the two-dimensional 
evolution
is given by the lowest allowed mode for fixed $m$. 

\begin{figure}[h]
\leavevmode
\epsfxsize=0.85\textwidth
\epsfbox{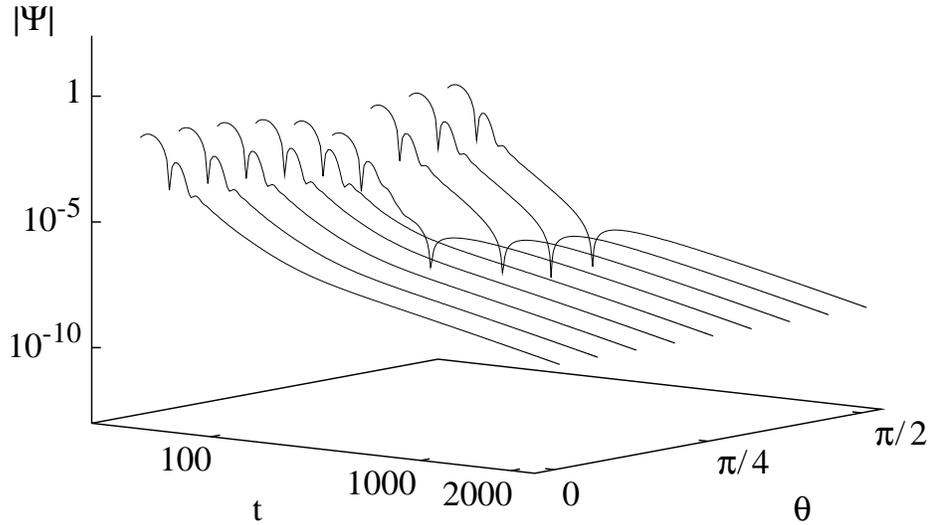}
\caption[figure6]{\label{figure6}
                 Log-log plots of
                 $|\Psi | = \mbox{Im}\Psi$ for $m=0$, $a=0.99$
                 at  $r^* = 10$ for different angular directions
                 in equidistant steps from $\theta = 0$ to
                 $\theta = \pi/2$.
                 The initial angular dependence was given by
                 $\mbox{Im}\Psi \sim P_2^0 \sim 3\cos^2\theta-1$.
                 Mixing of modes:
                 For large times the time dependence is given by
                 $| \Psi | \sim t^{-\mu}$, where for 
                 $\mbox{Im}\Psi$ the
                 exponent is $ 2.88 \leq \mu \leq 2.92$      with
$<\mu>= 2.91 $ over the observed angles. $\mbox{Re}\Psi=0$ at all
times, because the two equations are decoupled for $m=0$, and
$\mbox{Re}\Psi = \frac{d}{dt}\mbox{Re}\Psi= 0$ initially.
                       }
\end{figure}

The situation is different for the case $l=2$, $m=0$, depicted in
Fig.\ \ref{figure6}. Here the initial pulse is not given by the 
lowest
mode with $m=0$ that is symmetric with respect to the equatorial
plane, corresponding to $l=0$. Instead, the initial angular 
dependence
was given by $\mbox{Im}\Psi \sim P_2^0 \sim 3\cos^2\theta-1$.  
Here
mixing of modes occurs and the late-time evolution is dictated 
by the
lowest mode. The sink on the right-hand side is caused by the
transition to the lowest mode.  For large times the time 
dependence is
given by $| \Psi | \sim t^{-\mu}$, where $ 2.88 \leq \mu \leq 
2.92$
with $<\mu>= 2.91 $ when averaged over all observed angles.
In contrast to this result, the corresponding Schwarzschild case
exhibits no mixing of the modes, and
the theoretical power-law tail exponent is given by $\mu=7$ 
for $l=2,m=0$

\section{Acknowledgments}

We thank R.\ Gleiser, K.\ Kokkotas, H.-P.\ Nollert, J.\ Pullin, 
R.\
Price, and E.\ Seidel 
for helpful discussions.
This work was
supported by the Binary Black Hole Grand Challenge Alliance, NSF
PHY/ASC 9318152 (ARPA supplemented) and by NSF grants PHY 
93-09834,
93-57219 (NYI) to P.L.

\appendix

\section{}\label{appendix}

The horizon behavior of the wave equation written in 
Boyer-Lindquist coordinates is
illustrated here, in the context of the slow rotation 
approximation.
The manifestation of this behavior in the solution of the initial
value problem is analysed in the limit $r \rightarrow 2 M$.

The small $a$ expansion of equation (\ref{BL}) using the ansatz
\begin{equation}
\label{wrong1dans}
\Phi=\frac{1}{r}\,Y^m_l(\theta,\phi)\,[\Psi_{0}(t,r^*)+a\,
\hat\Psi_{1}(t,r^*)]\; ,
\end{equation}
yields the system of equations
\begin{mathletters}
\label{wrongsys}
\begin{eqnarray}
\label{wrongsys0}
\left[ -\partial_{tt} + \partial_{r^*r^*} - V(r) \right] \, 
\Psi_{0}
&&= 0 \, \\
\label{wrongsys1}
\left[ -\partial_{tt} + \partial_{r^*r^*} - V(r) \right] \, 
\hat\Psi_{1}
&&=  i\frac{4mM}{r^3}\,\partial_{t}\Psi_{0}  \, .
\end{eqnarray}
\end{mathletters}

Upon examining equation (\ref{wrongsys1}) a problematic feature
emerges.  The source appearing on the right hand side does not 
vanish in
the limit $r^* \rightarrow - \infty$. More precisely, the 
asymptotic 
form of the equation at the horizon is $(- \partial_{tt} + 
\partial_{r^*
r^*})\hat\Psi_1=\frac{i\,m}{2\,M^2}\,\partial_t \Psi_0$. Using null 
coordinates $v = t + r^{*}$ and $u = t - r^{*}$, and assuming a
purely ingoing solution $\Psi_{0}(v)$, one obtains,
\begin{equation}
\label{horizon}
\hat\Psi_{1}(u,v) =  b(u) \Psi_{0}(v)  + f_{1}(v) + f_{2}(u)\, ,
\end{equation}
where
\begin{equation}
b(u) = - \frac{i m}{8 M^2} u \, ,
\end{equation}
and $f_{1},f_{2}$ are solutions of the homogeneous equation.

The presence of this growing mode in the system may at first appear
puzzling, since the governing equations (\ref{BL}) and 
(\ref{teuks0})
are related simply by an angular coordinate redefinition. Yet the
singular nature of the transformation (\ref{kerrphi}) near the 
horizon
is the source of the growing mode.  A direct comparison of systems
(\ref{wrongsys}) and (\ref{numsys}) reveals that the two sets of
equations are equivalent, with $\hat\Psi_{1}$ given by
\begin{equation}
\label{linear}
\hat\Psi_{1}(t,r^{*}) = \Psi_{1}(t,r^{*}) + c(r)  
\Psi_{0}(t,r^{*})\, ,
\label{null}
\end{equation}
where
\begin{equation}
c(r) = \frac{\pi m}{2 M} - \frac{i m}{2 M} \ln(\frac{r}{r - 
2 M})\, .
\end{equation}
The solution of the equation with the correct source given by
(\ref{source})
is denoted by $\Psi_{1}(t,r^{*})$.
The coefficient $c(r)$ is singular at the horizon, hence 
bounded solutions $\Psi_{0}, \Psi_{1}$ give rise to an unbound
combination for $\hat\Psi_{1}$. The linear combination 
(\ref{linear})
has indeed the horizon singularity demonstrated by 
(\ref{horizon}). This singularity follows
from the fact that
\begin{equation}
\lim_{r \rightarrow 2M} c(r) = 
\lim_{r \rightarrow 2M} b(u) 
\approx  \lim_{r \rightarrow 2M} \frac{i m}{2M} \ln(r - 2 M)\, .
\end{equation}
An intuitive explanation of this pathology lies in the twisting 
of
azimuthal directions defined at infinity, as seen by an infalling
observer near the horizon. The wave evolution is well behaved in
local frame, yet it appears singular in a coordinate system that 
winds itself infinitely many times around the black
hole. The singularity of the evolution in the Boyer-Lindquist
coordinates is introduced entirely through this singular (at the
horizon) coordinate transformation.

%
%

%
%

%
%

%
%
%

%
%

\end{document}